\pdfoutput=1
\documentclass[iop]{emulateapj}

\usepackage{mathptmx,natbib,graphicx,grffile,wasysym,hyperref}

\newcommand{\he}{\rm He}
\newcommand{\hy}{\rm H}
\newcommand{\hi}{\rm H~\textsc{i}}
\newcommand{\hei}{\rm He~\textsc{i}}
\newcommand{\heii}{\rm He~\textsc{ii}}
\newcommand{\half}{\rm H$\alpha$}
\newcommand{\cm}{\rm cm$^{-2}$}
\newcommand{\cc}{\rm cm$^{-3}$}

\shorttitle{Polychromatic opacity imaging of densities}
\shortauthors{Williams, Baker \& van Driel-Gesztelyi}

\begin{document}

%%%%%%%%%%%%%%%%%%%%%%%%%%%%%%%%%%%%%%%%%%%%%%
\title{Mass estimates of rapidly-moving prominence material from high-cadence EUV images}
%%%%%%%%%%%%%%%%%%%%%%%%%%%%%%%%%%%%%%%%%%%%%%

\author{David~R.~Williams, Deborah~Baker and Lidia~van~Driel-Gesztelyi\altaffilmark{1,2}}
\affil{Mullard~Space~Science~Laboratory, University~College~London, Holmbury~St~Mary, Surrey, RH5~6NT, United~Kingdom.}
\email{d.r.williams@ucl.ac.uk}

\altaffiltext{1}{LESIA-Observatoire de Paris, CNRS, UPMC Univ. Paris 06, Univ.
Paris-Diderot, 92195 Meudon, France.}
\altaffiltext{2}{Konkoly Observatory, Budapest, Hungary}

%%%%%%%%%%%%%%%%%%%%%%%%%%%%%%%%%%%%%%%%%%%%%%
\begin{abstract}
We present a new method for determining the column density of erupting filament material using state-of-the-art multi-wavelength imaging data. Much of the prior work on filament/prominence structure can be divided between studies that use a polychromatic approach with targeted campaign observations, and those that use synoptic observations, frequently in only one or two wavelengths. The superior time resolution, sensitivity and near-synchronicity of data from the {\em Solar Dynamics Observatory}'s Advanced Imaging Assembly allow us to combine these two techniques using photo-ionisation continuum opacity to determine the spatial distribution of hydrogen in filament material. We apply the combined techniques to {\em SDO}/AIA observations of a filament which erupted during the spectacular coronal mass ejection on 2011 June 07. The resulting ``polychromatic opacity imaging'' method offers a powerful way to track partially ionised gas as it erupts through the solar atmosphere on a regular basis, without the need for co-ordinated observations, thereby readily offering regular, realistic mass-distribution estimates for models of these erupting structures.
\end{abstract}
%%%%%%%%%%%%%%%%%%%%%%%%%%%%%%%%%%%%%%%%%%%%%%

\keywords{methods: data analysis Ð radiative transfer Ð Sun: activity Ð Sun: coronal mass ejections (CMEs) Ð Sun: filaments, prominences Ð Sun: UV radiation}

%%%%%%%%%%%%%%%%%%%%%%%%%%%%%%%%%%%%%%%%%%%%%%
\section{Introduction}
%%%%%%%%%%%%%%%%%%%%%%%%%%%%%%%%%%%%%%%%%%%%%%
\label{intro}

%%%%%%%%%%%%%%%%%%%%%%%%%%%%%%%%%%%%%%%%%%%%%%%

The mass and structure of prominences/filaments are of interest in several aspects of the study of the corona. Since the launch of {\sl Hinode} and its Solar Optical Telescope \citep[SOT;][]{hinode,Tsuneta:2008p5422}, observations of the dynamic structures in prominences with SOT data \citep{Berger:2011p12846} have underlined the difficulties of explaining prominence dynamics. Development of radiative transfer models \citep[e.g.][]{Labrosse:2004p15399,Anzer:2005p12163} requires an understanding of the fine-scale mass structure and radiation field, while oscillations in prominences can be used for diagnosis of the magnetic field strength only if their mass can be accurately estimated \citep[and references therein]{Oliver:2009p15692}.  Models that seek to replicate the initial conditions of filament eruptions \citep{Low:2003p14968}, and those that address the ensuing propagation into the heliosphere \citep{DeForest:2012p15718} must take into account the effects of gravity \citep{Spicer:2006p14982} and composition \citep{Kilper:2009p14973} on these concentrations of material that are at least an order of magnitude denser than the surrounding corona \cite[e.g.,][]{Bommier:1994p1010,Jejcic:2009p15431}.

%%%%%%%%%%%%%%%%%%%%%%%%%%%%%%%%%%%%%%%%%%%%%%%
The most established methods of measuring mass distributions use Thomson scattering of photospheric light by free coronal electrons \citep{Hundhausen:1994p20195} to estimate the total amount of mass in prominences and (interplanetary) coronal mass ejections \citep[(I)CMEs;][and references therein]{Vourlidas:2006p15040}. More recently, investigators have used the amount of material evacuated in the dimming of the EUV corona to estimate the amount of mass lost to a CME/ICME \citep{Harrison:2000p15429,Aschwanden:2009p14986}, showing consistency with parallel observations from Thomson scattering estimates. 

At wavelengths shorter than the Lyman series limit (912~{\AA}), cool plasma is optically thick due to  photo-ionisation, principally of neutral hydrogen and neutral or singly-ionised helium. The product of abundance and cross-section for photo-ionisation of other species is negligible compared with these three. This continuum absorption is used in studies of the interstellar medium \citep[see][and references therein]{Vennes:1993p14737,Fruscione:1994p14664} to calculate the column density ($N_{i\;J}$) of these species along the line of sight, where $i$ is the element, and $J$ the ionisation stage ($J\in\{I, II, \cdots, Z\}$), since the total opacity at a given wavelength, $\tau(\lambda)$ is given by
\begin{equation}
\tau(\lambda) =	\sum_{i} \sum_J N_{ij} \sigma_{iJ}(\lambda),\\%\ .
\end{equation}
where $\sigma_{iJ}$ is the cross-section for absorption due to photo-ionisation for species $iJ$. This allows the column mass $M$ along the line of sight to be estimated as:
\begin{equation}
M = \sum_i m_i \sum_J N_{iJ}\ .\label{masseq}
\end{equation}
The effect was also noted in the solar transition region by \cite{Schmahl:1979p14871} as evidence of \hi\ and \hei\ -- and therefore partial ionisation -- in the chromosphere and transition region. \citep[See also the work of][]{Orrall:1976p14874,Kanno:1979p14965,Kanno:1982p14932}.

The most frequent investigations of column-density structure on the Sun are those that estimate how much light is attenuated by a filament, -- i.e., its opacity -- and, so, total mass along the line of sight in each spatial resolution element (Equation~\ref{masseq}). Since both band-pass imager and slit-spectrometer data of the Sun  exist in the EUV, these investigations have consisted of various combinations of images at a single wavelength or wavelength band \citep{Gilbert:2006p14739}, images at multiple wavelengths \citep{Golub:1999p12807,Engvold:2001p12809}, spectral line measurements in the EUV and (often) {\half} data \citep[{\it e.g.},][]{Schmieder:1999p14875,Mein:2001p14876}. 

\citet{Gilbert:2005p12806,Gilbert:2006p14739} have developed and applied a technique for estimating the mass of limb-crossing prominences using EUV images at 195~{\AA} only. This so-called ``spatial-interpolative'' approach estimates the background emission and prominence opacity. An interesting point is that they make use of the fact that {\em erupting} prominences eventually reveal something close to the original background emission, albeit at a later point in time. This allows a ``temporal-interpolative'' approach, to which we return later. Though not {\em strictly} ``monochromatic'', we use the term in this article to refer to all single-wavelength/-passband methods.

A logical extension of the above method is to use the information from several EUV wavelengths, over a sufficiently broad range, so that the cross-section for photo-absorption varies while other variables are assumed to hold constant. If all the associated radiation escapes from the corona, where the pressure scale-heights are large, then this approximation can be made.
Such ``polychromatic'' approaches include: those which are (coronal) image-based \citep{Golub:1999p12807,Gilbert:2011p12812}; those that use coronal line spectroscopy \citep{Kucera:1998p14382,Anzer:2005p12163}; and those that include emission from the transition region and/or chromosphere \citep[e.g.,][]{Schmieder:1999p14875,Mein:2001p14876}, often interpreted with the aid of non-LTE radiative transfer modelling to gain further insight into the detailed ionisation and mass structure of a filament. Motivated to constrain filament models, for example, \cite{Anzer:2003p14858} and \cite{Schwartz:2004p12164} used a radiative model \citep{Heinzel:2003p12165} to infer the height of filament material above the solar surface, and to differentiate between morphological filament models. The third of the listed techniques is still developing. \cite{Labrosse:2011p12160} most recently applied it to {\sl Hinode} EUV Imaging Spectrometer spectra \citep[EIS;][]{eis}, while \cite{Gilbert:2011p12812} exploited {\sl SoHO} imaging and spectroscopic data \citep{Domingo:1995p14832}. We refer the reader to the recent review by \cite{Labrosse:2010p15437} for an excellent treatment of these and other prominence diagnostics.

A distinct disadvantage of the spectroscopic approaches, pragmatically speaking, is that contemporary EUV spectrometers have a limited field of view, so that the observations must be targetted. {\half} measurements, on the other hand, are typically ground-based and so are susceptible to weather-imposed limitations. As a result, these methods have not been used on a regular basis.

A persistent unknown in much of this work is the background radiation field, here denoted $I_b$, that is attenuated by the filament material. The background corona is clearly variable on short and long spatial and temporal scales, but a view of it is blocked by the material under investigation. This has been treated by interpolating between points in the background on either side of cool masses \citep{Kucera:1998p14382}, with additional consideration of the profile of background coronal emission off-limb \citep{Gilbert:2011p12812,Golub:1999p12807}. However, because of the highly structured nature of the solar atmosphere, a better background model can be obtained by observing coronal emission in the absence of the absorbing structure, such as before and/or after a filament eruption. \cite{Gilbert:2005p12806} used both a ``temporal-interpolative'' approach and a ``spatial-interpolative'' approach to measure the column depth of a prominence which erupted from the south-east solar limb; their method was later applied to a larger sample of erupting and non-erupting prominences \citep{Gilbert:2006p14739}. Similarly, \cite{Kucera:2008p12845} use a temporal-interpolative approach in analysis of data from an erupting prominence structure taken with the SOHO/SUMER spectrometer \citep{Wilhelm:1995p14771}. 

It is worth pointing out that photometric measurements are complicated by the presence of filament cavities \citep{Gibson:2010p10296} where, for coronal lines, there is lower emissivity in both the cavity and the filament itself. This effect is referred to as ``volume blocking'' or ``emissivity blocking'' \citep{Heinzel:2003p12165}, and simply refers to a lack of emission measure. The amount of attenuated radiation emerging from behind the filament is difficult to disentangle from radiation emitted in the foreground, but a promising alternative to the interpolative approaches is to use the X-ray transparency of filament material \citep{Anzer:2007p12243} and assume that the EUV background emission can be scaled to match the spatial soft X-ray distribution \citep{Heinzel:2008p12162}. Whilst this, too, is an assumption, it does address the issue of reduced emission in the cavity.

Since the launch of NASA's {\it Solar Dynamics Observatory} \citep{Pesnell:2011p14740} incorporating the Advanced Imaging Assembly \citep[AIA;][]{Lemen:2011p13102}, we have access to much higher-cadence observations of the solar corona, with data simultaneous or near-simultaneous in several EUV wavelengths. This allows us to apply a temporal-interpolative approach to measure the total background and foreground emission around cool-material structures in apparent motion across the Sun, often at several points close in time to the transient absorption taking place. 

%%%%%%%%%%%%%%%%%%%%%%%%%%%%%%%%%%%%%%%%%%%%%%%
In this article, we describe the application of two methods -- one monochromatic, the other polychromatic -- to AIA observations of cool material returning from a spectacular filament eruption. Although the eruption on 2011 June 7 is successful, producing a CME, some filament material returns to the solar atmosphere and surface, and our aim is to infer the density of material from its obscuration of the background corona. We describe two approaches to achieving this aim. The first approach calculates the opacity of the material from data taken in a single AIA filter, an approach also used in \cite{projectlidia} but described here in detail. The second approach uses arguments first made by \cite{Kucera:1998p14382} to separate the column density from the product of filling factor and geometrical depth using observations made at multiple wavelengths. In both cases, we use properties of the opacity due to {\hi}, {\hei} and {\heii} photo-ionisation continua to give a more powerful lower-limit estimate of the total hydrogen column density along the line of sight, using the vastly improved cadence and sensitivity of AIA data. Lastly, we compare the results of these approaches.

%%%%%%%%%%%%%%%%%%%%%%%%%%%%%%%%%%%%%%%%%%%%%%%
\section{Observations}
%%%%%%%%%%%%%%%%%%%%%%%%%%%%%%%%%%%%%%%%%%%%%%%
\label{obs}

The observations used in this work were recorded by AIA on 2011 June 7, in the period 06:30 -- 08:00~UT. The region of interest is around NOAA active region 11226, which produced an M2.7-class X-ray flare and a spectacular filament eruption, in which most of the filament mass appears to have returned to the Sun. The returning filament mass is the subject of this study, where we estimate the column and volumetric densities. We focus on two primary targets in which to diagnose the total hydrogen column density, $N_{\hy}$. 

\begin{figure*}
\begin{center}
	\includegraphics[width=5.3 in,bb=18 0 507 243]{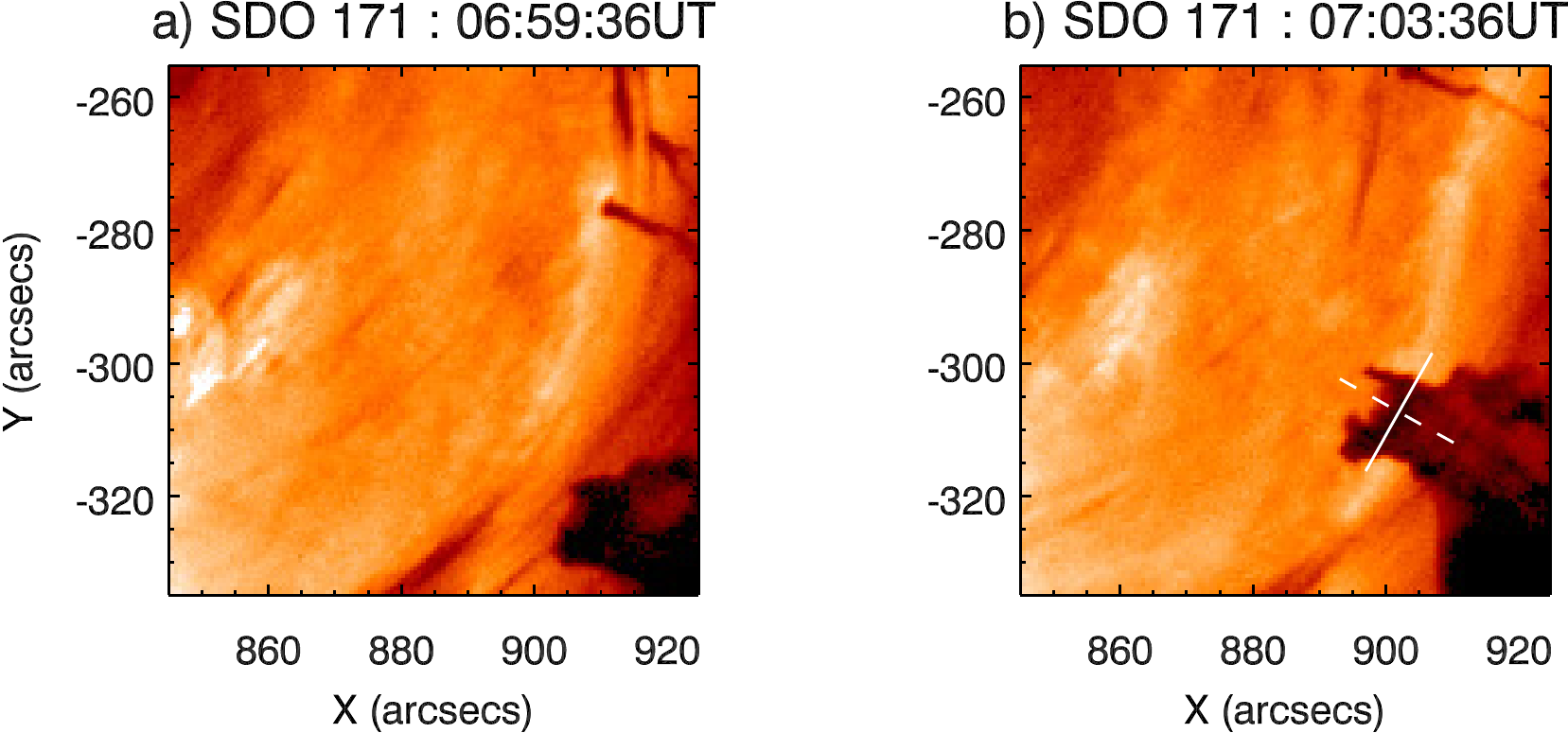}
\caption{a) Portion of an AIA 171~{\AA} filter image which is used as the model unattenuated image ($I_0$) for Target~1, a concentration of filament material that has failed to escape during the CME under study. b) Target~1 itself, which has now fallen further towards the Sun. Dashed and solid intersecting lines refer to the positions of profiles taken through $N_{\hy}$ shown in Figure~\ref{slicethrublob}. (See the electronic edition of the Journal for a color version of this figure, where it is also available as an MPEG
animation.)}
\label{sliceblob}
\end{center}
\end{figure*}

Target~1 is a large concentration of returning material that appears to fall unhindered by solar magnetic field, suggesting high plasma $\beta$, consistent with a low ionisation degree. The material is shown in Figure~\ref{sliceblob}b. 

Target~2 is an area to the east of the active region, where returning filament mass arrives at what resembles a Y-shaped coronal magnetic null-point \citep[Figure~\ref{rxnh}; see also][]{projectlidia}. Material that appears dark in the EUV range enters and leaves this area and we estimate the density of one of the departing concentrations.

In order to correctly estimate the measurement errors in our data, we model the photon noise using the calibration curves distributed in the AIA branch of the {\it SolarSoft} IDL software library \citep[SSW;][]{Freeland:1998p493}.

%+-+-+-+-+-+-+-+-+-+-+-+-+-+-+-+-+-+-+-+-+-+-+-+-+-+-+-+-+-+-+
%+-+-+-+-+-+-+-+-+-+-+-+-+-+-+-+-+-+-+-+-+-+-+-+-+-+-+-+-+-+-+

\section{Method}
\label{meths}
In this article, we assume that the dominant process which removes photons from the line of sight is photo-ionisation. We construct the cross-sections for photo-ionisation, $\sigma_{i\;J}(\lambda)$, using the analytical approximations given by \cite{Verner:1996p12166}. We have also considered the effect of \hei\ autoionisation resonances, using Fano profile parameters given by \cite{Rumph:1994p12156}. Although $\sigma_{\hei}$ can be enhanced by as much as a factor of 10 in these resonances, the effect is very narrow-band (full width at half-maximum $\lesssim 0.5$~{\AA}) when compared with the width of the AIA bandpasses (FWHM of AIA 193 channel = 6.3~{\AA}), so we disregard the effect of these resonances in calculating \hei\ opacity. Additionally, we assume that there is no emission from the prominence material in the wavebands observed (we return to this issue briefly in Section~\ref{disco}).

\begin{figure}
\begin{center}
\includegraphics[height=3.7in,angle=0]{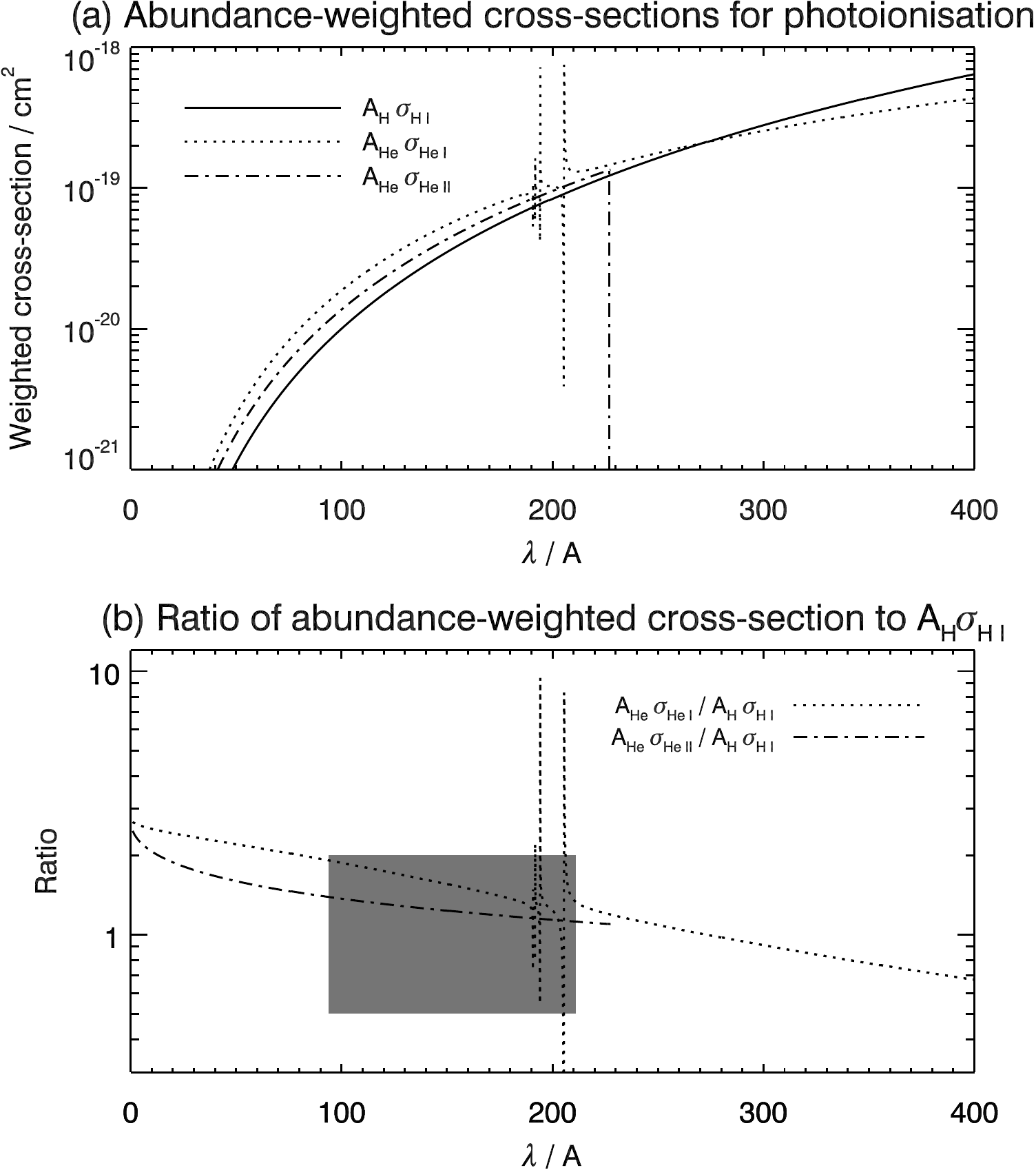}
\caption{(a) cross-sections for photo-ionisation, $A_i \sigma_{iJ}$, for the three principal contributors in the short EUV; (b) ratio of $A_i \sigma_{iJ}$ to $A_{\rm H}\sigma_{\hi}$. The grey shaded box indicates (horizontally) the range of EUV wavelengths shorter than the \heii\ ionisation edge that are observed by AIA, as well as the range within which the abundance-weighted cross-section can deviate by a factor of $\leq 2$ from that of \hi\ (vertically). \label{figweighted}}
\end{center}
\end{figure}

As noted by \cite{Daw:1995p12811}, $\sigma_{\hei}$ and $\sigma_{\heii}$ are similar in value (for $\lambda \leq 227$~\AA). Here, however, we draw attention to the fact that the cross sections of the first three species are very similar when weighted by elemental abundance, {\em i.e.}, $A_i \sigma_{iJ}$ (Figure~\ref{figweighted}). We follow the convention that $A_{\hy}$ is unity, and take the value $A_{\he} = 0.0851$ from \cite{Grevesse:2007p10802}. This lets us make the approximation
\begin{eqnarray}
\tau_{He} & = & N_{\hei}\,\sigma_{\hei} + N_{\heii}\,\sigma_{\heii} \nonumber \\
&  \approx  & (N_{\hei}+N_{\heii})\:\sigma_{\heii} \nonumber \\
&  \approx  & A_{\rm He}\,N_{\rm H}\,\sigma_{\heii}
\end{eqnarray}
and since $\tau_{\hi}(\lambda) \approx \tau_{\he}(\lambda)$, for $\lambda < 227~{\mathrm{\AA}}$, $\tau_{tot} \approx 2\tau_{\he}$. Allowing for the fact that some helium may be fully ionised, we can then estimate the total {\em hydrogen} column density along the line of sight as
\begin{equation}
N_{\hy} \ge \frac{\tau_{tot}}{2A_{\he}\sigma_{\heii}}\label{ourapprox}.
\end{equation}
The deviation of the abundance-weighted cross-sections of helium from that of \hi\ is, in fact, less than a factor of 2 (see shaded box in Figure~\ref{figweighted}b) for all the wavelengths of AIA data analysed here. Because the above approximation is expressed in terms of the helium cross-sections, it allows us to more accurately gauge the total hydrogen column density for weak degrees of {\heii} ionisation, independent of the ionisation degree of hydrogen, which fully ionises at temperatures much lower than does helium. Accordingly, and considering the inequality sign in Equation~\ref{ourapprox}, our estimate of $N_{\hy}$ will be a lower limit; the lower the ionisation degree of helium, the more this value is likely to represent an accurate value rather than just a lower limit.
%
%+-+-+-+-+-+-+-+-+-+-+-+-+-+-+-+-+-+-+-+-+-+-+-+-+-+-+-+-+-+-+
%+-+-+-+-+-+-+-+-+-+-+-+-+-+-+-+-+-+-+-+-+-+-+-+-+-+-+-+-+-+-+
%

%%%%%%%%%%%%%%%%%%%%%%%%%%%%%%%%%%%%%%%%%%%%%%%
\subsection{Polychromatic measurements}
%%%%%%%%%%%%%%%%%%%%%%%%%%%%%%%%%%%%%%%%%%%%%%%
\label{multiple}
\cite{Kucera:1998p14382} write the following expressions for the observed EUV intensity at a given point on a filament, considering the effects of a pixel filling factor along the line of sight, $f$:
%
%+-+-+-+-+-+-+-+-+-+-+-+-+-+-+-+-+-+-+-+-+-+-+-+-+-+-+-+-+-+-+
\begin{eqnarray}
I_{obs} 	& = & I_b \left [f e^{-\tau} + (1 - f) \right] + I_f \nonumber \\
%              	& = & (I_b + I_f)  - fI_b + f e^{-\tau}I_b \nonumber \\
              	& = & I_0 - fI_b(1 - e^{-\tau}\label{tau1})
\end{eqnarray}
%+-+-+-+-+-+-+-+-+-+-+-+-+-+-+-+-+-+-+-+-+-+-+-+-+-+-+-+-+-+-+
where
%
%+-+-+-+-+-+-+-+-+-+-+-+-+-+-+-+-+-+-+-+-+-+-+-+-+-+-+-+-+-+-+
\begin{equation}
I_0 = I_b + I_f .
\end{equation}
This allows Equation~\ref{tau1} to be rewritten as:
\begin{eqnarray}
1 - \frac{I_{obs}}{I_0}	& = & f\frac{I_b}{I_0}(1 - e^{-\tau})\label{obseq}\\
				& = & G(1 - e^{-\tau})\label{eqg}
\end{eqnarray}
%+-+-+-+-+-+-+-+-+-+-+-+-+-+-+-+-+-+-+-+-+-+-+-+-+-+-+-+-+-+-+
where the factor $G$ combines the two unknown geometrical factors of the fraction of emission behind the absorbing material, $I_b/I_0$ and the filling factor that describes how much of the pixel area is filled by this material. 
\cite{Anzer:2005p12163} note that the former term can potentially be appreciably less than unity, particularly when the absorbing material is low in the atmosphere. The quantity on the left-hand side of Equation~\ref{obseq} is an observable quantity, denoted the absorption depth, $d$.
%+-+-+-+-+-+-+-+-+-+-+-+-+-+-+-+-+-+-+-+-+-+-+-+-+-+-+-+-+-+-+
\begin{eqnarray}
d & \equiv & 1 - \frac{I_{obs}}{I_0}\label{zaobs}\\
d(\lambda) & = & G[1 - e^{-\tau(N_{\hy}\,;~\lambda)}]\label{zafunk}\\
\mathit{i.e.,} & & \nonumber \\
F(G, N_{\hy}\,;~\lambda) & = & d(\lambda) \label{jax}
\end{eqnarray}
%+-+-+-+-+-+-+-+-+-+-+-+-+-+-+-+-+-+-+-+-+-+-+-+-+-+-+-+-+-+-+
In this case, $d(\lambda)$ can be fitted to a function  $F$ if there are independent observations at a sufficient number of wavelengths.

The above method is appropriate where we have simultaneous observations of the same target. This also holds true for {\em near}-simultaneous observations, provided that the target does not change shape or density appreciably between measurements. Thanks to the cadence of AIA, this is true in the case of Target~1. 
We therefore cross-correlate images of monochromatic opacity, $\tau(\lambda)$ (see Section~\ref{monoc}) to ensure accurate co-registration, since this property ought to scale largely with $\tau(\lambda)$ between measurements at different wavelengths.

\begin{figure}
\begin{center}
\hspace{-0.5cm}
\includegraphics[width=7cm]{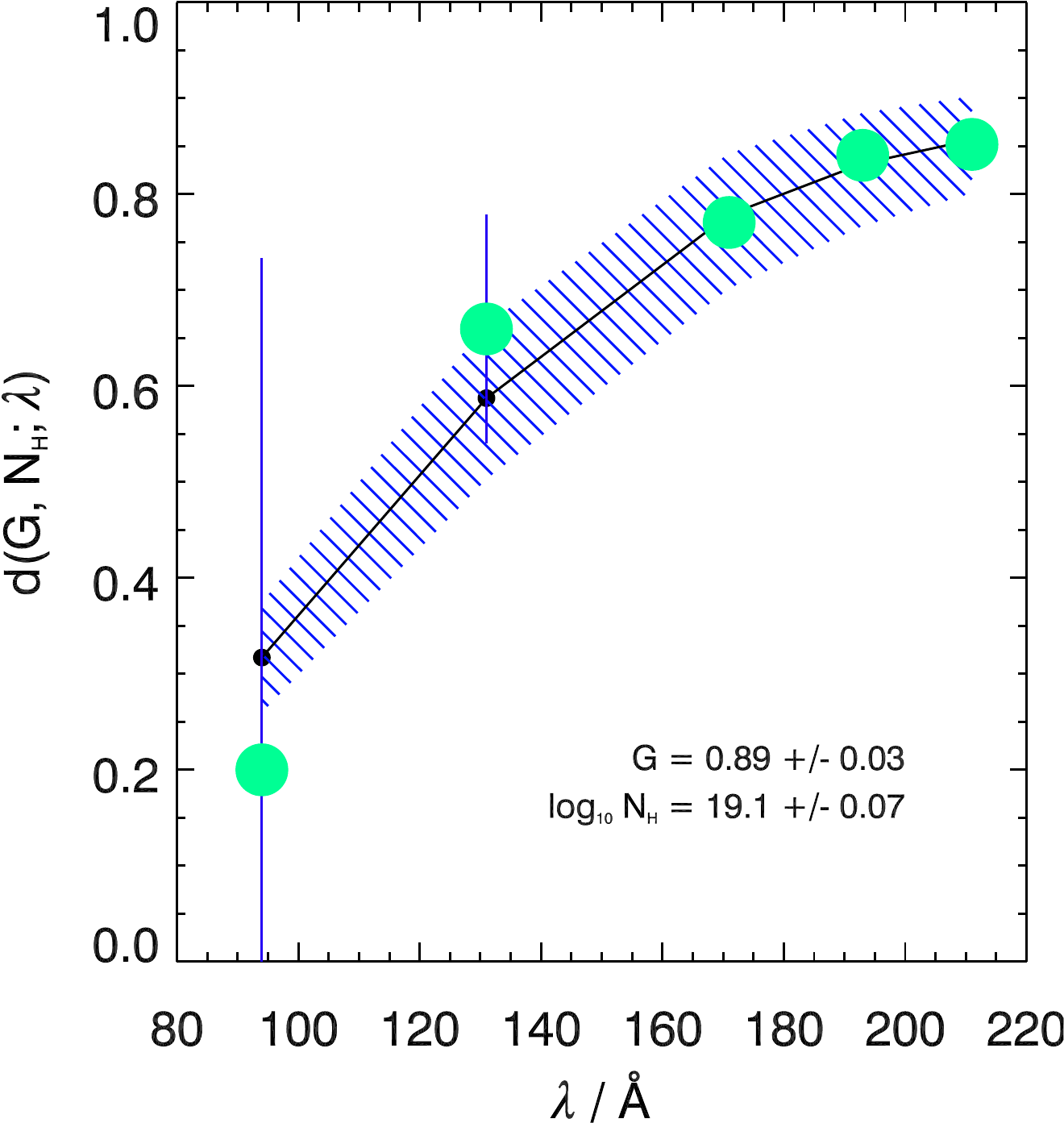}
\caption{Example of the result of fitting observed $d(\lambda)$ at all AIA wavelengths below 227~{\AA}. Data points are taken from a measurement inside the large mass shown in Figure~\ref{sliceblob}b. Large circles with error bars: $d(\lambda)$. Small circles joined by solid line: values of $d$ due to the best fit to Equation~\ref{zafunk}, where the hatched envelope represents the bounds of the best fit $\pm 1\sigma$. (See the electronic edition of the Journal for a color version of this figure.)}
\label{sample}
\end{center}
\end{figure}

In this analysis, we use a Levenberg-Marquardt least-squares minimisation algorithm to find the best fit to the measured $d(\lambda)$ in Equation~\ref{zafunk}. An example fit is shown in Figure~\ref{sample}. Since there are two free parameters of $F$, we require observations in at least three wavelengths to constrain them. We use data from the 94, 131, 171, 193, and 211~{\AA} channels. Although the images are near-synchronous rather than being truly cotemporal, the cadence of the observations is such that the form of Target~1 does not change appreciably between images at these wavelengths. We disregard data from the 304~{\AA} channel as this emission is optically thick, while we are unable to use data from the only other EUV channel (335~{\AA}) because of what appear to be stray- or scattered-light effects, exacerbated by the extremely bright flare emission from the active region where this eruption originates. The remaining AIA filters measure wavelengths that are longer than the Lyman series limit. Figure~\ref{NHmap} shows the resulting best-fit values of $N_{\hy}$ and $G$ for Target~1.

A difficulty in selecting the background area for this target is that for most of the duration of this event, many other concentrations of erupted filament material cross the field of view in various directions, and only in the images taken around 06:59:36~UT is there a clear view of the solar background emission -- strictly, $I_0$ -- without attenuation. We therefore choose frames at (or nearest to) this time as our estimate of $I_0(\lambda)$.

%%%%%%%%%%%%%%%%%%%%%%%%%%%%%%%%%%%%%%%%%%%%%%%
\subsection{Opacity measurements at a single wavelength}
%%%%%%%%%%%%%%%%%%%%%%%%%%%%%%%%%%%%%%%%%%%%%%%
\label{monoc}
The filament mass concentrations in Target~2 move, and deform, much more rapidly than Target~1, so that it is not possible to convincingly co-register images taken at different times and wavelengths. This being the case, we cannot use near-simultaneous measurements of intensity reduction at different wavelengths to fit Equation~\ref{zafunk}. However, since we can no longer estimate $G$ independently of $N_H$, we can simply set $G = 1$ and accept that we will make an underestimate of the column density in this way, since this implies a unity filling factor, and that all EUV emission $I_0$ is behind the erupting filament material. In this case, we would rewrite Equation~\ref{zafunk} for a single wavelength as simply
\begin{eqnarray}
\tau({\lambda}) & \geq & \ln I_{0,\lambda} - \ln I_{obs,\lambda}.
\end{eqnarray}
Using this monochromatic opacity, we are again able to calculate a map of the lower limit to this quantity, and of the resulting column density show in Figures~\ref{NHmap} \& \ref{rxnh}c . Note that this type of opacity map is used co-register the near-simultaneous images at different wavelengths in Section~\ref{multiple}.

The area covered by the field of view in Figure~\ref{rxnh} is, fortunately, unobscured by returning filament material until around 06:50~UT. There is a pronounced dimming around this region following the transit of a coronal wave, but the period between 06:44 and 06:48~UT lies between the passage of the wave and the arrival of falling filament material. Therefore, we model the background emission, $I_0$, as the mean intensity of images taken during this interval (Figure~\ref{rxnh}a). Because we calculate $I_0$, in this case, from a time-average of these data, we estimate the measurement error on the background at each pixel as the quadrature sum of the photon noise from each image and the standard deviation of intensity in that pixel in time. %{\em express mathematically?}

%%%%%%%%%%%%%%%%%%%%%%%%%%%%%%%%%%%%%%%%%%%%%%%
\section{Analysis}
%%%%%%%%%%%%%%%%%%%%%%%%%%%%%%%%%%%%%%%%%%%%%%%

%%%%%%%%%%%%%%%%%%%%%%%%%%%%%%%%%%%%%%%%%%%%%%%
\subsection{Polychromatic method}
%%%%%%%%%%%%%%%%%%%%%%%%%%%%%%%%%%%%%%%%%%%%%%%
\label{polyc}

\begin{figure*}
\begin{center}
\includegraphics[width=6.5cm,angle=90]{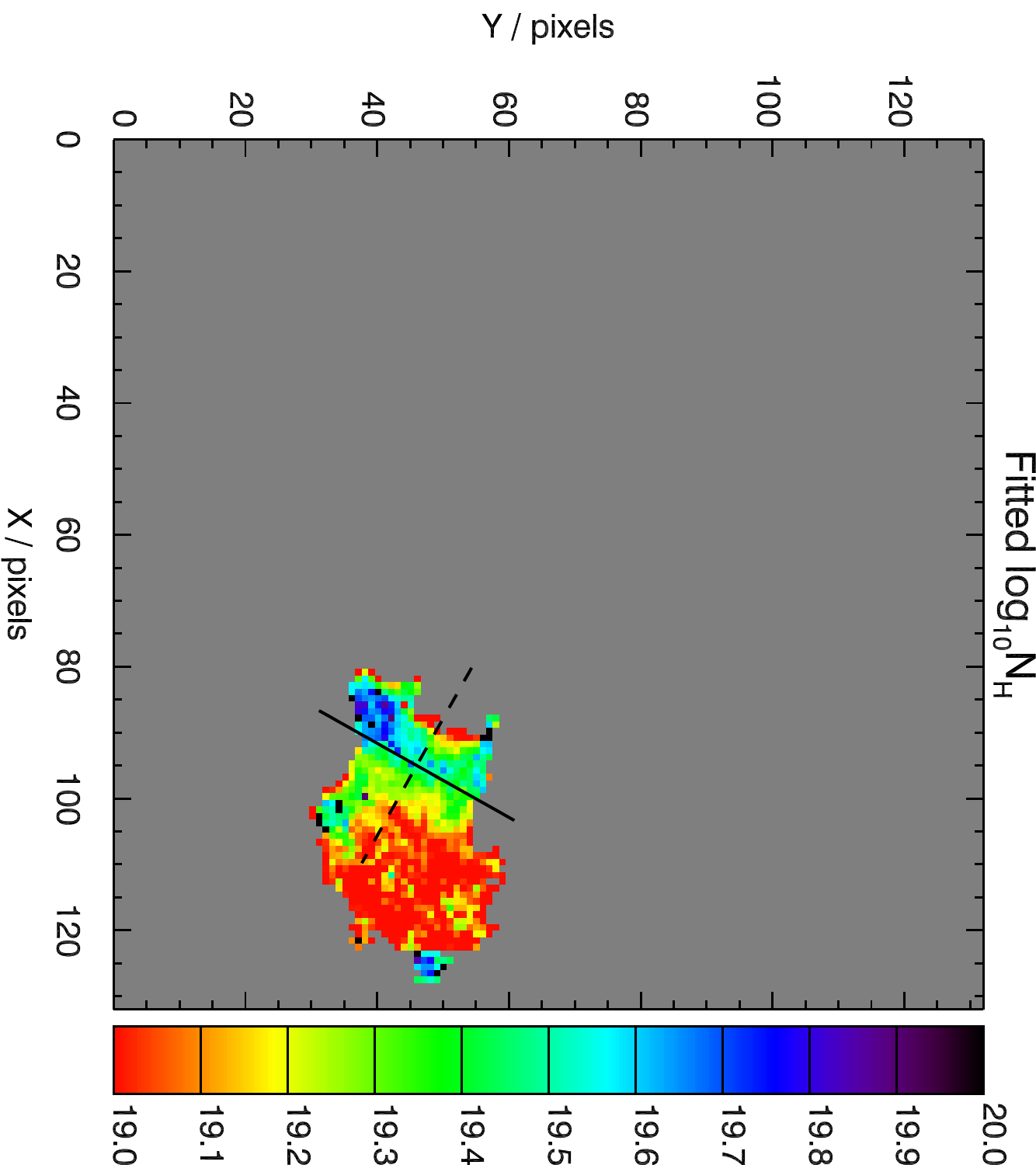}
\hspace{1cm}
\includegraphics[width=6.5cm,angle=90]{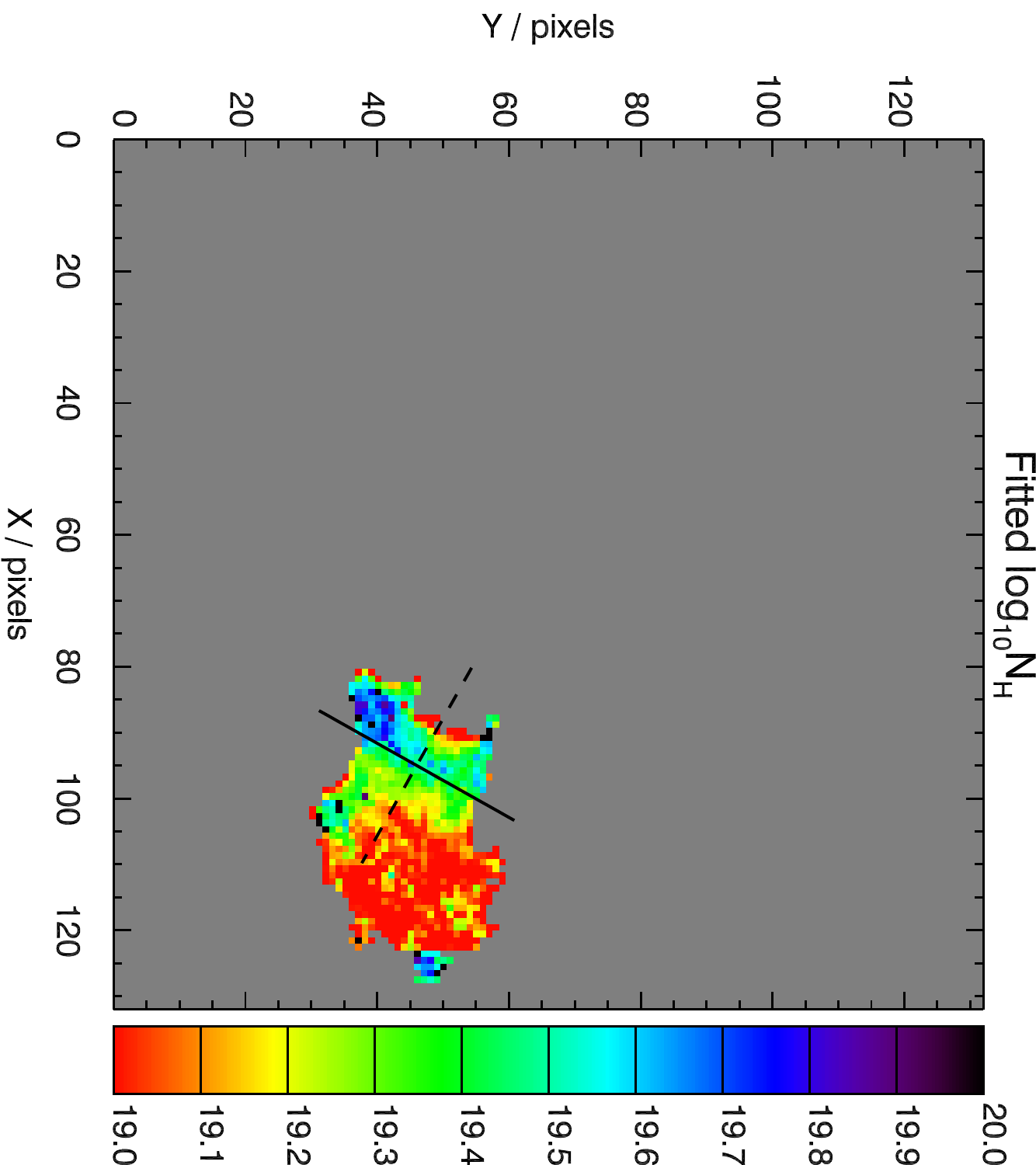}
\caption{Left: Best-fit column density map for hydrogen, obtained by fitting measured absorption depth $d$ (Equation~\ref{zaobs}), to the function $F$ (eqs.~\ref{zafunk} \& \ref{jax}). Right: as left, but showing the best-fit map of geometric factor $G = f I_b/I_0$\label{NHmap}. The field of view shown is identical to that shown in Figure~\ref{sliceblob}. (See the electronic edition of the Journal for a color version of this figure.)}
\end{center}
\end{figure*}

We define the edge of Target~1 as the level where $G = 0.6$ (Figure~\ref{NHmap}). This level is set arbitrarily, but seems to match the visual edge of the target well in the portion of the FOV which was unobscured in $I_0$. The restriction of our analysis to the area indicated in Figure~\ref{NHmap} reflects the fact that some portions of the image used as $I_0$ were already obscured by absorbing material. The animation that accompanies Figure~\ref{NHmap} in the online edition of the journal shows that other, fainter filament material moves between the times indicated between Figures~\ref{NHmap}a (constituting $I_0$) and ~\ref{NHmap}b ($I_{obs}$), revealing an elongated bright structure behind Target~1. This would tend to cause an underestimate of $N_{\hy}$ in the affected pixels, since the bright structure was not accounted for in $I_0$, so that the drop in intensity would in fact be slightly larger.

\begin{figure}
\begin{center}
	\includegraphics[width=3.3in]{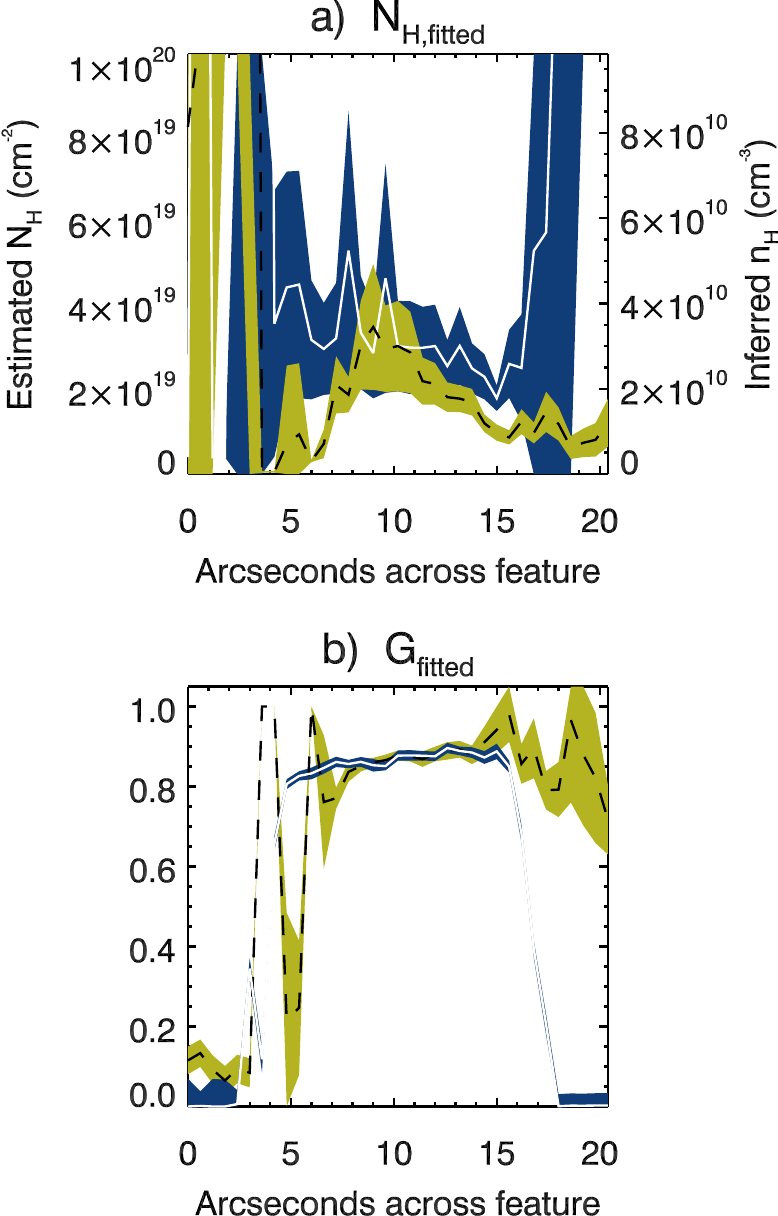}
\caption{a) Total hydrogen column density, and inferred volumetric density, along the slices through Target~1 shown in Figure~\ref{sliceblob}b. b) Along the same slices, the value of the geometric factor, $G$ is plotted. For both figures, the dashed line corresponds to the dashed slice in Figure~\ref{sliceblob}b, while the solid line corresponds to the solid slice (See the electronic edition of the Journal for a color version of this figure.)}
\label{slicethrublob}
\end{center}
\end{figure}

The detailed map of best-fit column density (Figure~\ref{NHmap}a) shows a concentration of material towards the lowest part of the target, with column densities of $N_{\hy} \sim 10^{20}$~\cm. We note that these values are rather larger than those previously reported in either erupting or quiescent filaments \citep{Penn:2000p15528,Gilbert:2005p12806,Labrosse:2010p15437}. The numeric values and their associated $1\sigma$ errors along two slices are shown in Figure~\ref{slicethrublob}. While the variance along a slice in $N_{\hy}$ is noticeable, the corresponding profiles of $G$ are much less variable. In fact, the value of $G$ seems remarkably stable across the target: given that Target~1 is likely to have reasonably uniform $I_b / I_0$ on larger scales, its small-scale uniformity indicates that the pixel filling factor, $f$ may be rather smooth. $G$ also lies close to unity, which indicates that $f$, too, must be rather high. The absolute value of $I_b/I_0$ is tricky to determine (without using stereoscopic data), but the high value of $G$ indicates that the mass must be several emission scale heights into the corona.  

We should point out that the values along the slices include {\em all} the values fitted in the FOV shown in Figure~\ref{NHmap}, not only those within the target shown, which leads to the more wildly varying fitted values -- from outside the $G = 0.6$ contour -- being plotted at the edges of Figure~\ref{slicethrublob}a.

In order to estimate mean volumetric (number) densities in this structure, we divide $N_{\hy}$ by a characteristic path length through the material. We choose the FWHM of $G$ (14'') as our indicator, equivalent to a path length of $l = 10^9$~cm. The estimates of $n_{\hy}$ are therefore dependent on the value of $l$ being constant; although this is almost certainly an oversimplification for any given point in the target, it serves as a guide for estimating the total hydrogen density, which reaches $n_{\hy} \gtrsim 10^{10}$~\cc.

%+-+-+-+-+-+-+-+-+-+-+-+-+-+-+-+-+-+-+-+-+-+-+-+-+-+-+-+-+-+-+
%+-+-+-+-+-+-+-+-+-+-+-+-+-+-+-+-+-+-+-+-+-+-+-+-+-+-+-+-+-+-+

%+-+-+-+-+-+-+-+-+-+-+-+-+-+-+-+-+-+-+-+-+-+-+-+-+-+-+-+-+-+-+
%+-+-+-+-+-+-+-+-+-+-+-+-+-+-+-+-+-+-+-+-+-+-+-+-+-+-+-+-+-+-+

%+-+-+-+-+-+-+-+-+-+-+-+-+-+-+-+-+-+-+-+-+-+-+-+-+-+-+-+-+-+-+
%+-+-+-+-+-+-+-+-+-+-+-+-+-+-+-+-+-+-+-+-+-+-+-+-+-+-+-+-+-+-+

%%%%%%%%%%%%%%%%%%%%%%%%%%%%%%%%%%%%%%%%%%%%%%%
\subsection{Monochromatic method}
\begin{figure*}
\begin{center}
	\includegraphics[width=6.8in]{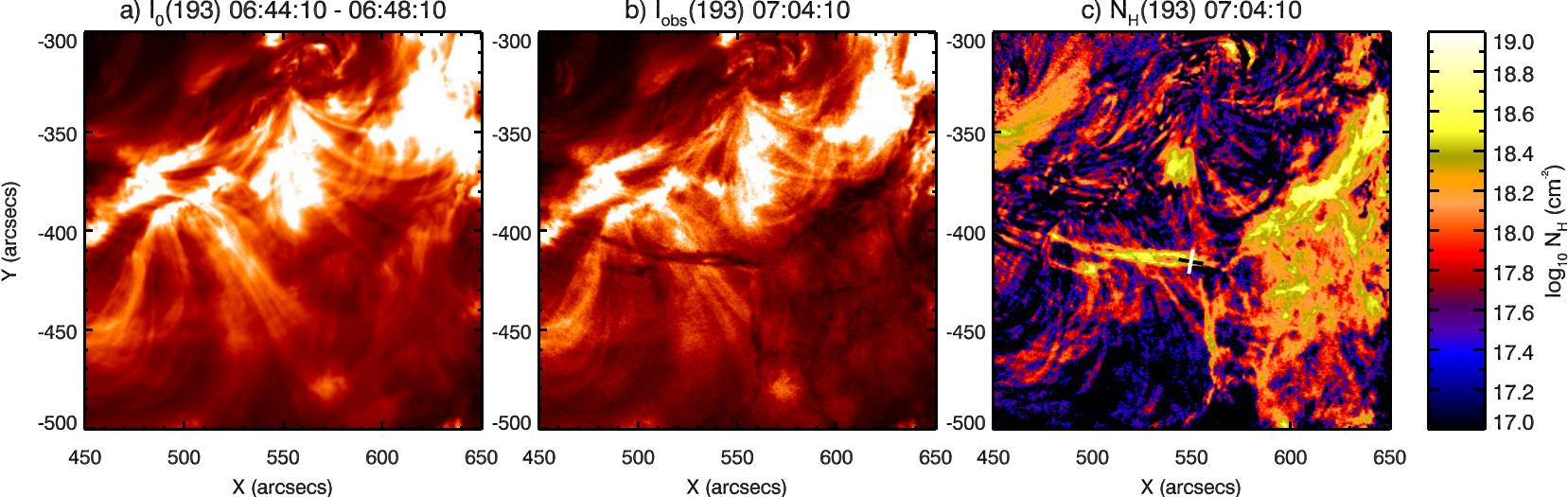}
\caption{a) Time average of AIA 193~{\AA} data between the times indicated above, representing the model of unattenuated intensity from the corona, $I_0$. b) Image from the same channel of AIA showing in-falling filament material that reaches what appears to be a magnetic null point, near position (560'', -410''), at which the material is subsequently deflected east and west. In this image, the background radiation is attenuated in places by the filament material, and so represents $I_{obs}$. c) Column density of hydrogen, $N_{\hy}$, estimated with the monochromatic method described in Section~\ref{monoc}. The intersecting black and white lines at (550'', -410'') indicate the position of profiles of $n_{\hy}$, shown in Figure~\ref{rxslices}. (See the electronic edition of the Journal for a color version of this figure, where it is also available as an MPEG animation.)}
\label{rxnh}
\end{center}
\end{figure*}

\begin{figure}
\begin{center}
	\includegraphics[width=3.3in]{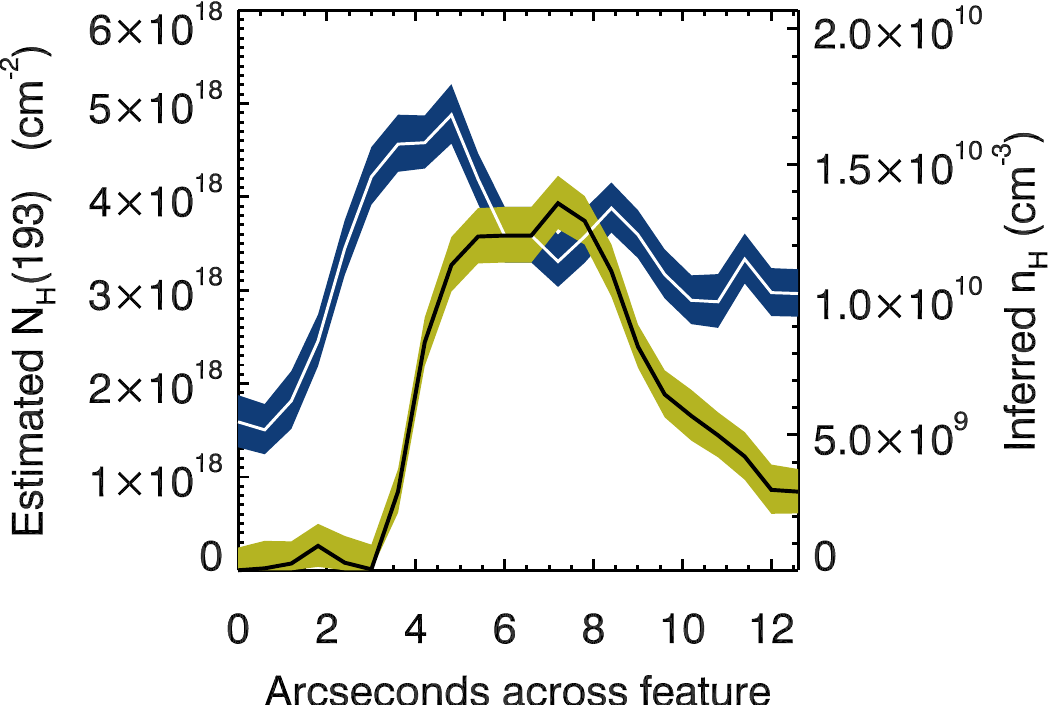}
	%\vspace{2in}
\caption{Hydrogen column density, and inferred volumetric electron density, along the slices through Target~2: the black/white line with $1\sigma$ envelope corresponds to the black/white profile position shown in Figure~\ref{rxnh}c. (See the electronic edition of the Journal for a color version of this figure.)}
\label{rxslices}
\end{center}
\end{figure}

%%%%%%%%%%%%%%%%%%%%%%%%%%%%%%%%%%%%%%%%%%%%%%%
\label{workingout}
As stated in \S~\ref{monoc}, the primary object to which we apply our monochromatic method is Target~2 (Figure~\ref{rxnh}b). In this case, we have used data from the AIA 193~{\AA} channel because of its superior signal-to-noise ratio. In contrast to our approach in \S~\ref{polyc}, we do not isolate any one piece of material in the reduced field-of-view shown. However, we once again take profiles of $N_{\hy}$ through a concentration of absorbing material, seen to be moving eastward (negative $x$ direction) from the apparent Y-point. The positions of these profiles are indicated in Figure~\ref{rxnh}c, and the corresponding values along each are shown in Figure~\ref{rxslices}. It is noticeable that $N_{\hy}$ varies more smoothly across this target than in Target~1 (Figure~\ref{slicethrublob}). Although the nature of the monochromatic analysis is a mapping from a more smoothly-varying intensity to $\tau$ (rather than a fit of two free parameters to multiple data points), the background being time-averaged (\S~\ref{monoc}) may also contribute to this smoothness.

More generally, the amount of material contained in these thinner `threads' is still substantial, with lower limit values in the range $N_{\hy} \gtrsim 10^{18}$ -- $10^{19}$~{\cm}, similar to those seen in more extended prominences \citep{Kucera:1998p14382}.
In order to estimate the volumetric density $n_{\hy}$, we again assume a representative path length equivalent to the FWHM of the structure in the plane-of-sky, which we estimate from Figure~\ref{rxslices} as 4'' ($3\times 10^8$~cm). Again, this is an oversimplification, but it suggests values of $n_{\hy} \gtrsim 10^{10}$~{\cc} within this target filament mass,  comparable with those found for the much larger Target~1.

%+-+-+-+-+-+-+-+-+-+-+-+-+-+-+-+-+-+-+-+-+-+-+-+-+-+-+-+-+-+-+
%+-+-+-+-+-+-+-+-+-+-+-+-+-+-+-+-+-+-+-+-+-+-+-+-+-+-+-+-+-+-+
%
%+-+-+-+-+-+-+-+-+-+-+-+-+-+-+-+-+-+-+-+-+-+-+-+-+-+-+-+-+-+-+
%+-+-+-+-+-+-+-+-+-+-+-+-+-+-+-+-+-+-+-+-+-+-+-+-+-+-+-+-+-+-+

%%%%%%%%%%%%%%%%%%%%%%%%%%%%%%%%%%%%%%%%%%%%%%%
\subsection{Comparison of methods}
%%%%%%%%%%%%%%%%%%%%%%%%%%%%%%%%%%%%%%%%%%%%%%%
Since the monochromatic method is more appropriate to fast-changing/-moving features, where we cannot co-register images taken at different times, it is of interest to gauge how well this method performs when compared with our polychromatic approach (Section~\ref{multiple}) for Target~1. 

\begin{figure}
\begin{center}
	\includegraphics[width=8cm]{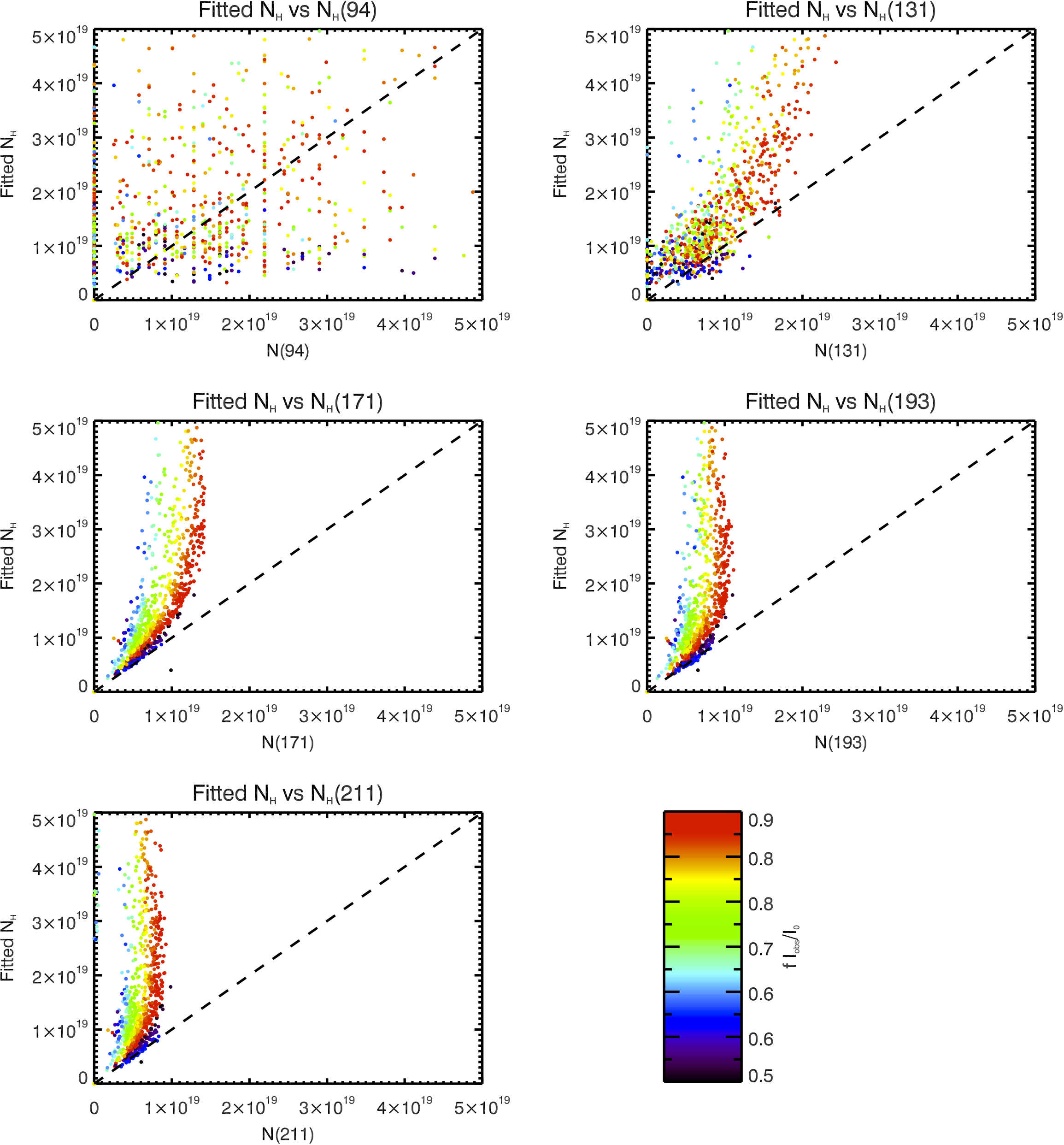}
\caption{Correlation plots of the best-fit H column depth, on the horizontal axis, against that derived from monochromatic absorption. Colour-coding represents the corresponding geometric factor, $G$ for each data point. With the exception of $N_{\hy}(94)$ and $N_{\hy}(131)$, the two lowest signal-to-noise channels, the monochromatic result is consistently an underestimate when compared to the fitted $N_{\hy}$. (See the electronic edition of the Journal for a color version of this figure.)}
\label{dotcompare}
\end{center}
\end{figure}

Figure~\ref{dotcompare} shows correlation plots of monochromatic $N_{\hy}(\lambda)$ to polychromatic $N_{{\hy}, fitted}$.
In the two shortest-wavelength channels, the signal to noise is rather poor, and the scatter in correlation is large. However, for the 171, 193 and 211~{\AA} channels, the scatter is less, and $N_{\hy}(\lambda)$ is shown to be consistently underestimated when compared with $N_{\hy, fitted}$. 
In the same figure, we have colour-coded the points to show the fitted value of $G$; in general, the smaller the value of $G$, the greater the underestimate of the column density when only a single wavelength is used. This is perhaps unsurprising, since we set the filling factor arbitrarily to unity in calculating the monochromatic values.

%+-+-+-+-+-+-+-+-+-+-+-+-+-+-+-+-+-+-+-+-+-+-+-+-+-+-+-+-+-+-+
\begin{figure}
\begin{center}
	\includegraphics[width=8cm]{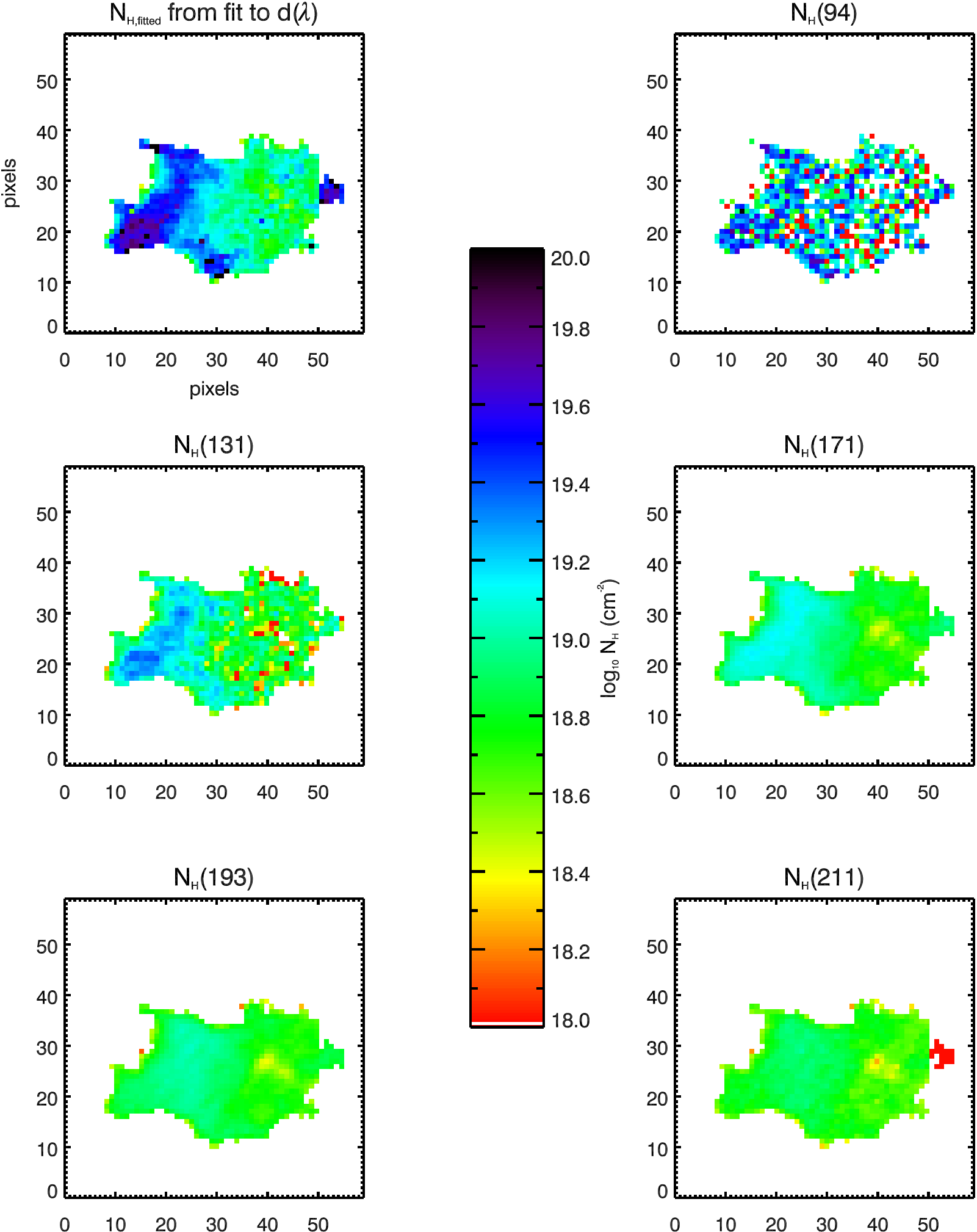}
\caption{Comparison of the best-fit H column depth in Target~1, derived polychromatially (top left), with that derived from individual wavelength opacities (remaining plots). Values are plotted logarithmically according to the colour scale in the centre. (See the electronic edition of the Journal for a color version of this figure.)}
\label{multicomparison}
\end{center}
\end{figure}
%+-+-+-+-+-+-+-+-+-+-+-+-+-+-+-+-+-+-+-+-+-+-+-+-+-+-+-+-+-+-+

%+-+-+-+-+-+-+-+-+-+-+-+-+-+-+-+-+-+-+-+-+-+-+-+-+-+-+-+-+-+-+
%+-+-+-+-+-+-+-+-+-+-+-+-+-+-+-+-+-+-+-+-+-+-+-+-+-+-+-+-+-+-+

%%%%%%%%%%%%%%%%%%%%%%%%%%%%%%%%%%%%%%%%%%%%%%%
\section{Discussion}
%%%%%%%%%%%%%%%%%%%%%%%%%%%%%%%%%%%%%%%%%%%%%%%
\label{disco}

Each of our approaches to determining $N_{\hy}$ depends upon determining its value through the expression in Equation~\ref{ourapprox}. We assume a fixed relative abundance for helium (Section~\ref{meths}), but it should be noted that \cite{Gilbert:2007p15401} and \cite{Kilper:2009p14973} find variation of the He/H abundance ratio in prominences, pointing to cross-field diffusion of neutral H and He as the most likely cause of stratification of this ratio. However, their finding is unlikely to apply since the dynamic timescale in the targets analysed here is much shorter than the diffusion timescale \citep{Gilbert:2007p15401}. In eruptive events, \cite{Kilper:2009p14973} also find that the ratio tends to homogenise in the hours leading up to eruption, and we suggest that the mixing of plasma induced by the eruption is likely to contribute further to this homegenisation. 
Work by \cite{Bemporad:2009p15441} uses the {\sl Hinode} EUV Imaging Spectrometer \citep{hinode,eis} to show significant, spatially unresolved velocities ($\xi \lesssim 100$~km~s$^{-1}$) at the location of an erupting prominence, which may support the picture of mixing. 

Since the absorbers of EUV radiation we consider here are ions and neutrals in a partially ionised plasma, we can only measure the total column density of neutral and partially ionised helium (thus, hydrogen), and estimate $n_{\hy}$;  we can make no comment on the density of free electrons in this plasma beyond the trivial $0 < n_e < (n_{\hy}/0.83)$, since the ratio $n_e / n_{\hy}$ will depend upon the detailed parameters of the plasma and radiation field. Any fixing of the electron-to-hydrogen ratio to the usual coronal value of 0.83 is likely to lead to overestimation of $n_e$ in the targets considered here. 

Similarly, we are unable to make estimates of the ionisation degree of helium, since: 1) as we have shown, the cross-section for absorption of {\hei} is very close to that of {\heii} (Section~\ref{meths}); and 2) we have no useful data longward of the \heii\ ionisation edge (227~{\AA}; \S~\ref{multiple}). Nor do we have data in both the helium continua and the \hi\ Lyman continuum ($504 < \lambda < 912$~{\AA}), which precludes measurements of He/H abundance variations.

We also do not consider the information in the distribution of $N_{\hy}$ about how thick Target~1 might be at each pixel -- the concentration of mass in the lower parts is ignored, along with information that that may impart. The method can of course be inverted, assuming a constant value for $n_{\hy}$ to infer a spatially varying path length through the material.

The approaches used in this article lack an advantage of the spectroscopic approaches favoured by previous authors. For example, we do not analyse emission from individual emission lines, some with very short scale height, to make estimates of the effect of lowered emission due to the cavity around a filament as \cite{Heinzel:2003p12165} do. However, in the case of an erupting filament, the cavity rapidly expands, so that both $I_0$ and $I_f$ are sampled within the cavity at the time of our measurements. 
Our use of emission from filters rather than at individual wavelengths is not a difficulty, however, since the cross-sections for photo-ionisation depend only on wavelength.

An earlier, spectroscopic analogue of the work by \cite{Kucera:1998p14382} applied to erupting prominences can be found in \cite{Penn:2000p15528}, who uses five emission lines between 555 and 630~{\AA} to derive column densities of around $5\times 10^{17}$~{\cm}. Although there are already likely to be significant differences between the amount of material studied by \citeauthor{Penn:2000p15528} and that studied here, they also use lines in a region of the spectrum which has been shown by \cite{Mein:2001p14876} and \cite{Gilbert:2011p12812} to suffer from saturation of the photo-absorption effect. The much higher column densities we find may be due to the lack of these saturation effects at the shorter wavelengths studied here.

%%%%%%%%%%%%%%%%%%%%%%%%%%%%%%%%%%%%%%%%%%%%%%%
\subsection{Underestimation factors}
%%%%%%%%%%%%%%%%%%%%%%%%%%%%%%%%%%%%%%%%%%%%%%%

In Figures~\ref{slicethrublob} and \ref{rxnh}, we infer the volumetric electron density $n_e$ on the right-hand axis of the plotted profiles. In each case, we infer the density by estimating a path depth equivalent to the width of the feature observed. For the profiles in Target~1, we estimate a path length of 14'' ($\sim 10^9$~cm); for the narrower feature in Target~2, we estimate a path length of 4'' ($\sim 3\times10^8$~cm). These depths are estimated from the FWHM of the corresponding $N_{\hy}$ profiles.

Studying the \hei~D$_3$ and {\half} lines in a number of prominences, \cite{Bommier:1994p1010} find mean volumetric electron densities of $(2.1 \pm 0.7) \times 10^{10}$~cm$^{-3}$. The values we infer from a monochromatic measurement of our targets are consistent with this mean, if a little on the low side. However, the monochromatic estimates are systematically lower than those derived from the multi-wavelength fit to $d(\lambda)$ by as much as an order of magnitude.
In addition to this effect, there is the consideration that our estimates are again under-representations of the true column depth if a non-vanishing fraction of the helium along the line of sight is fully ionised, a distinct possibility if heating takes place in or around this material. In an erupting structure, \cite{Kucera:2008p12845} find evidence of emission at $\log T_e \sim 5.0$. 

As already indicated above, the column depth can become so large that it effectively blocks all emission from behind. \cite{Gilbert:2011p12812} recently encountered problems with the $\lambda$625 line, finding that the esimate of $N_{\hy}$ at $\lambda$195 was much larger. The saturation of the effect at longer $\lambda$  will cause  the depth to be underestimated: at some limiting value of $\tau$, no additional photons can be scattered/absorbed by the absorbing mass. Further evidence of this effect is shown by \cite{Heinzel:2008p12162} as disagreement between estimates of the continuum opacity derived at Mg~\textsc{x} $\lambda$625 with that inferred from $\tau_{\hy\alpha}$. The former shows an underestimate of column density compared with estimates from the latter. \cite{Gilbert:2011p12812}  Each paper attributes the problem of underestimated mass \citep[as do][]{Kucera:1998p14382} to the saturation effect described above.

These effects would suggest that the ratio between $N_H$ and the column depth of all absorbers deviates from the lower limit of Equation~\ref{ourapprox}. However, there is an additional term which may lead to our underestimation of $N_H$. As stated in Section~\ref{meths}, we do not consider the effects of emission from the prominence material in the EUV passbands analysed. There could, in fact be emission in lines over a wide range of temperatures; \cite{ODwyer:2010p10347} have recently performed a detailed analysis of the emission lines (and corresponding temperatures) to which AIA is sensitive. Differential emission measures constructed from sections of a prominence by \cite{Parenti:2007p12844} suggest that there is a considerable amount (though not a majority) of material at $\log T_e > 4.8$. Either of these cases would suggest that the effect on our results will be to make them a lower limit to the local filament mass, as already expected. Furthermore, in response to the work of \cite{Kucera:1998p14382}, \cite{Engvold:2001p12809} point out that the former's results are susceptible to emission from hot lines in and around filaments before eruption. They are unable to establish the contribution of these hot lines to the total observed emission, but filaments are certainly susceptible to heating before and during eruption, as shown by \cite{Kucera:2008p12845}, often with apparently coronal-temperature emission \cite[e.g.,][]{Liewer:2009p10813,Sterling:2011p16226} as is perhaps seen in Target~2 (Figure~\ref{rxnh}). It is less clear whether there is heating of Target~1, the target in which we apply \cite{Kucera:1998p14382}'s method, as it returns to the lower solar atmosphere, but this possibility is certainly worth bearing in mind. In any case, the measurements made here would still serve as a lower limit.

%%%%%%%%%%%%%%%%%%%%%%%%%%%%%%%%%%%%%%%%%%%%%%%
\subsection{Advantages to this method}
%%%%%%%%%%%%%%%%%%%%%%%%%%%%%%%%%%%%%%%%%%%%%%%

Despite the disadvantages mentioned above, our approaches have a number of advantages:
\begin{enumerate}
\item They require no co-ordinated observing campaigns, only cotemporal or near-cotemporal images of an erupting (or otherwise rapidly moving) filament in several EUV wavelengths. Full-sun EUV observations are carried out in a patrol like fashion all the time by AIA, so that the approaches can be applied whenever data are available.
\item Because we use lines in the range where the helium cross-sections are at least as important as that of hydrogen, the method is sensitive at higher temperatures than those which use the Lyman continuum only ($504 < \lambda < 912$~{\AA}).
\item In exploiting the similarity of the \hei\ and \heii\ photo-ionisation cross-sections, we are able to measure the presence of all but bare helium, which allows us to estimate the total hydrogen column density, not only that of neutral hydrogen, through an abundance argument (Equation~\ref{ourapprox}).
\item The images used here are co-temporal or nearly co-temporal, allowing us to combine EUV image data in a way which has previously only been possible with spatially rastering or fixed-slit spectrometer data.
\item \label{emblock} When applied to erupting filaments in an expanding cavity, the complication of emissivity blocking is removed, as the cavity rapidly expands to encompass the observed area including $I_0$ (but see below for a comment on this).
\item Material from an erupted filament is likely to also be several scale heights into the corona, meaning that the  majority of emission will be from behind the erupting material, as suggested by Figure~\ref{NHmap}.
\item The method is, in principle, also applicable in a spatial-interpolative way.
\end{enumerate}

A qualification of point~\ref{emblock} is that emissivity blocking can be understood as low-temperature material replacing higher-temperature plasma that would otherwise contribute to the emission seen by AIA. In this sense, it is not limited to prominence cavities, and could in fact lead to some over-estimation of the absorption due to the filament, since the absorption depth is a measure of the decrease in intensity. However, since the filament material we have studied is likely to be rather high in the atmosphere, this consideration is unlikely to be important here.

\cite{Heinzel:2008p12162} draw an interesting (but probably incorrect) distinction between two types of observation previously used to carry out this type of analysis: those typically only obtained at one wavelength by large-field of view instruments (``patrol-type''); and those that use a combination of as many spectral lines as possible, termed ``multi-wavelength'' observations, largely performed by a superset of the authors in \cite{Heinzel:2008p12162} and \cite{Kucera:1998p14382}. Perhaps because of the lack of co-temporal multi-wavelength data at the time, their definition of ``multi-wavelength'' refers to spectrometer observations, rather than images. We argue that the terminology is no longer valid since our method, which we term ``polychromatic opacity imaging'', uses the patrolling advantages of a full-time, full-Sun imager, in combination with the (near-)co-temporal images from multiple passbands in the photo-ionisation continuum to construct column density maps of large-scale features. Whereas contemporary EUV spectrometers are nearly always rastering,  the examples we present have spatial simultaneity, and can therefore be applied to large-field observations. 

The results of fitted $N_{\hy}$ shown in Figure~\ref{multicomparison} reveal a surprising amount of coherence in their detail, and this may reflect agreement with a prior test carried out by \cite{Gilbert:2005p12806}. These authors compared the results of spatial- and temporal-interpolative approaches where possible, and found that the temporal-interpolative approach yields more accurate (and precise) results when compared with Thomson-scattering estimates.

Lastly, our approaches have been restricted to narrower fields-of-view in this article, but this is largely driven by the desire to test the techniques. In principle, the methods can be applied to full-disc images. {\em SDO}/AIA is well suited to this task, and the current rise toward solar maximum would suggest many suitable targets for the method.

%%%%%%%%%%%%%%%%%%%%%%%%%%%%%%%%%%%%%%%%%%%%%%%
\section{Conclusions}
%%%%%%%%%%%%%%%%%%%%%%%%%%%%%%%%%%%%%%%%%%%%%%%

We present a study which shows the powerful application of polychromatic opacity imaging to cotemporal and near-cotemporal data from {\sl SDO}/AIA of a spectacular filament eruption. The single- and multi-passband approaches used in this study give consistent lower limits on the total hydrogen column density (as opposed to just the neutral hydrogen column density) in the temperature regime below full ionisation of helium, and therefore are sensitive to a large range in prominence temperature. The values of column hydrogen density that we find are larger than in previous studies, but one of the targets chosen is rather exceptional and large column masses are perhaps to be expected.

Since we restrict our study to estimating the column density and inferring the volumetric density of  hydrogen, this method removes the need for co-ordinated observing campaigns to study prominence mass through comparison of opacity at multiple wavelengths. Using AIA data allows us to perform such studies of erupting filaments wherever they are observed.

Such observational constraints may be useful in the future development of more realistic models of erupting filament mass in coronal mass ejections.

\acknowledgments

DRW acknowledges a UCL University Fellowship and DRW, DB and LvDG all acknowledge an STFC Rolling Grant, for the funding of this research. LvDG acknowledges funding through the Hungarian Science Foundation grant
OTKA K81421. The research leading to these results has received funding from the European Commission's Seventh Framework Programme under the grant agreement no. 284461 (eHEROES project). The authors also acknowledge the JHelioviewer team \citep{jhelioviewer} for enabling fast searching of the data used in this study. JHelioviewer is part of the ESA/NASA Helioviewer Project. DRW would like to thank Sarah Matthews, Lucie Green, David Brooks and Nicolas Labrosse for valuable discussion on aspects of this work, and the referee for thoughtful comments and advice on the article.

{\it Facilities:} \facility{SDO (AIA)}.

%\bibliographystyle{hapj}
%\bibliography{omni}

\end{document}